\input harvmac
\input graphicx

\def\Title#1#2{\rightline{#1}\ifx\answ\bigans\nopagenumbers\pageno0\vskip1in
\else\pageno1\vskip.8in\fi \centerline{\titlefont #2}\vskip .5in}
%

%
%
\ifx\includegraphics\UnDeFiNeD\message{(NO graphicx.tex, FIGURES WILL BE IGNORED)}
\def\figin#1{\vskip2in}
\else\message{(FIGURES WILL BE INCLUDED)}\def\figin#1{#1}
\fi
\def\Fig#1{Fig.~\the\figno\xdef#1{Fig.~\the\figno}\global\advance\figno
 by1}
%
%
%
%

%
%
\font\ticp=cmcsc10

\def\calo{{\cal O}}
\def\calh{{\cal H}}
\def\calhh{{\hat{\cal H}}}

\def\roughly#1{\mathrel{\raise.3ex\hbox{$#1$\kern-.75em\lower1ex\hbox{$\sim$}}}}

\def\ahat{{\hat a}}
\def\ahats{|\ahat\rangle}
\def\as{|a\rangle}

\def\rtt{{\sqrt2}}
\def\zhat{{\hat 0}}
\def\ohat{{\hat 1}}
\def\zhats{|\zhat\rangle}
\def\ohats{|\ohat\rangle}
\def\zs{|0\rangle}
\def\os{|1\rangle}
\def\qhat{{\hat q}}
\def\uhat{{\hat U}}
\overfullrule=0pt
%
%
\lref\Hawkrad{
  S.~W.~Hawking,
  ``Particle Creation By Black Holes,''
  Commun.\ Math.\ Phys.\  {\bf 43}, 199 (1975)
  [Erratum-ibid.\  {\bf 46}, 206 (1976)].
}
\lref\PageUP{
  D.~N.~Page,
  ``Black hole information,''
[hep-th/9305040].
}
\lref\SGTrieste{S.~B.~Giddings,
  ``Quantum mechanics of black holes,''
  arXiv:hep-th/9412138
  }
\lref\SGinfo{S.~B.~Giddings,
    ``The black hole information paradox,''
  arXiv:hep-th/9508151.
}
\lref\Mathurrev{
  S.~D.~Mathur,
  ``The Information paradox: A pedagogical introduction,''
Class.\ Quant.\ Grav.\  {\bf 26}, 224001 (2009).
[arXiv:0909.1038 [hep-th]];  ``What the information paradox is {\it not},''  
  [arXiv:1108.0302 [hep-th]].
}
\lref\BHMR{
  S.~B.~Giddings,
  ``Black holes and massive remnants,''
Phys.\ Rev.\  {\bf D46}, 1347-1352 (1992).
[hep-th/9203059].
}
\lref\thooholo{
  G.~'t Hooft,
  ``Dimensional reduction in quantum gravity,''
  arXiv:gr-qc/9310026.
}
\lref\sussholo{
  L.~Susskind,
  ``The World As A Hologram,''
  J.\ Math.\ Phys.\  {\bf 36}, 6377 (1995)
  [arXiv:hep-th/9409089].
}
\lref\LQGST{
  S.~B.~Giddings,
  ``Locality in quantum gravity and string theory,''
Phys.\ Rev.\  {\bf D74}, 106006 (2006).
[hep-th/0604072].
}
\lref\NLvC{
  S.~B.~Giddings,
  ``Nonlocality versus complementarity: A Conservative approach to the information problem,''
Class.\ Quant.\ Grav.\  {\bf 28}, 025002 (2011).
[arXiv:0911.3395 [hep-th]].
}
\lref\Page{
  D.~N.~Page,
  ``Information in black hole radiation,''
  Phys.\ Rev.\ Lett.\  {\bf 71}, 3743 (1993)
  [arXiv:hep-th/9306083].
}
\lref\HaPr{
  P.~Hayden, J.~Preskill,
  ``Black holes as mirrors: Quantum information in random subsystems,''
JHEP {\bf 0709}, 120 (2007).
[arXiv:0708.4025 [hep-th]].
}
\lref\Braunstein{
  S.~L.~Braunstein and K.~Zyczkowski,
  ``Entangled black holes as ciphers of hidden information,''
[arXiv:0907.1190 [quant-ph]]; S.~L.~Braunstein, M.~K.~Patra,
  ``Black hole evaporation rates without spacetime,''
Phys.\ Rev.\ Lett.\  {\bf 107}, 071302 (2011).
[arXiv:1102.2326 [quant-ph]].
}
\lref\Bekenstein{
  J.~D.~Bekenstein,
  ``Black holes and entropy,''
Phys.\ Rev.\  {\bf D7}, 2333-2346 (1973).
}
\lref\QBHB{
  S.~B.~Giddings,
  ``Quantization in black hole backgrounds,''
  Phys.\ Rev.\  D {\bf 76}, 064027 (2007)
  [arXiv:hep-th/0703116].
}
\lref\GiLi{
  S.~B.~Giddings, M.~Lippert,
  ``The Information paradox and the locality bound,''
Phys.\ Rev.\  {\bf D69}, 124019 (2004).
[hep-th/0402073].
}
\lref\HoMa{
  G.~T.~Horowitz, J.~M.~Maldacena,
  ``The black hole final state,''
JHEP {\bf 0402}, 008 (2004).
[arXiv:hep-th/0310281 [hep-th]].
}
\lref\Rozali{
  B.~Czech, K.~Larjo, M.~Rozali,
  ``Black Holes as Rubik's Cubes,''
[arXiv:1106.5229 [hep-th]].
}
\lref\Sussrem{
  L.~Susskind,
  ``Trouble for remnants,''
[hep-th/9501106].
}
\lref\wabhip{
  S.~B.~Giddings,
  ``Why aren't black holes infinitely produced?,''
Phys.\ Rev.\  {\bf D51}, 6860-6869 (1995).
[hep-th/9412159].
}
\lref\SeSu{
  Y.~Sekino, L.~Susskind,
  ``Fast Scramblers,''
JHEP {\bf 0810}, 065 (2008).
[arXiv:0808.2096 [hep-th]].
}
\lref\GiSh{S.~B.~ Giddings and Y.~ Shi, work in progress.}
\lref\BHIUN{
  S.~B.~Giddings,
  ``Black hole information, unitarity, and nonlocality,''
Phys.\ Rev.\  {\bf D74}, 106005 (2006).
[hep-th/0605196].
}
\lref\fuzz{
  S.~D.~Mathur,
  ``Fuzzballs and the information paradox: A Summary and conjectures,''
[arXiv:0810.4525 [hep-th]].
}

\lref\Hawkunc{
  S.~W.~Hawking,
  ``Breakdown Of Predictability In Gravitational Collapse,''
  Phys.\ Rev.\  D {\bf 14}, 2460 (1976).
}

\lref\BPS{
  T.~Banks, L.~Susskind and M.~E.~Peskin,
  ``Difficulties For The Evolution Of Pure States Into Mixed States,''
  Nucl.\ Phys.\  B {\bf 244}, 125 (1984).
}
\lref\Astrorev{
  A.~Strominger,
  ``Les Houches lectures on black holes,''
  arXiv:hep-th/9501071.
}
\lref\LPSTU{
  D.~A.~Lowe, J.~Polchinski, L.~Susskind, L.~Thorlacius and J.~Uglum,
  ``Black hole complementarity versus locality,''
  Phys.\ Rev.\  D {\bf 52}, 6997 (1995)
  [arXiv:hep-th/9506138].
}
\lref\GiNe{
  S.~B.~Giddings and W.~M.~Nelson,
  ``Quantum emission from two-dimensional black holes,''
  Phys.\ Rev.\  D {\bf 46}, 2486 (1992)
  [arXiv:hep-th/9204072].
}
\lref\BaFitherm{
  T.~Banks, W.~Fischler,
  ``Space-like Singularities and Thermalization,''
[hep-th/0606260].
}
\Title{
\vbox{\baselineskip12pt
}}
{\vbox{\centerline{Models for unitary black hole disintegration}
}}
\centerline{{\ticp 
Steven B. Giddings\footnote{$^\ast$}{Email address: giddings@physics.ucsb.edu}  
} }
\centerline{\sl Department of Physics}
\centerline{\sl University of California}
\centerline{\sl Santa Barbara, CA 93106}
\vskip.10in
\centerline{\bf Abstract}

Simple models for unitary black hole evolution are given in an effective Hilbert-space description, parameterizing a possible minimal relaxation of locality, with respect to semiclassical black hole geometry.
 
\vskip.3in
\Date{}

\newsec{Introduction}

The unitarity crisis or  black hole (BH) information problem challenges the pillars of modern physics:  quantum mechanics (QM), Lorentz/diffeomorphism invariance, and locality.  In short, Hawking's argument for BH evaporation\refs{\Hawkrad} yields information loss\refs{\Hawkunc}, and it has been argued that there is no consistent scenario with these pillars intact.\foot{For reviews see \refs{\PageUP\SGTrieste\Astrorev\SGinfo-\Mathurrev}.}

Locality, in particular, seems a weak link in quantum gravity, and there have been suggestions to modify it\refs{\BHMR\thooholo\sussholo\LQGST\BHIUN-\NLvC} in some way.  However, such modification should be subtle as locality is a basic principle of local quantum field theory (LQFT), which describes observed phenomena extremely well.  Generally, we observe that quantum information is well localized in spacetime, and moreover generic nonlocality leads to causality paradoxes.  

One thus seeks a consistent dynamics describing deeper relations between quantum information and spacetime and its symmetries, where locality may be approximate but not exact, yet consistent unitary evolution is intact. 

Assuming QM, a useful tool is an effective quantum information-theoretic parameterization, with a simple Hilbert-space description of the dynamics.  Such an approach has been used {\it e.g.} in \refs{\Page\QBHB\HaPr-\Braunstein,\Mathurrev,\Rozali}.  Approximate locality suggests we decompose the overall Hilbert space into a product of Hilbert spaces $\calhh$ and $\calh$ corresponding to states inside and outside a BH, with evolution providing specific ``small" couplings between these.  

Such a framework is general enough to describe local evolution, but also nonlocal modifications.  Our approach will seek a conservative alternative\refs{\NLvC} to complementarity/hol\-ography\refs{\thooholo,\sussholo}, staying as close as possible to LQFT, with minimal, controlled allowance for nonlocality, as needed for unitary evolution.  In particular, in the evolution of \Hawkrad, $\calhh$ is effectively infinite dimensional, but unitarity and other indicators suggest an effective number of BH degrees of freedom $N(M) = \log({\rm dim} \calhh)$ at BH mass $M$ that is finite, {\it e.g.}  given by the Bekenstein-Hawking\refs{\Bekenstein,\Hawkrad} entropy,
\eqn\ndof{N(M)=S_{BH}(M)\ .}
A problem is how to describe physical unitary evolution incorporating such finite information content for a BH.

We will begin by reviewing the familiar scenario of Hawking evaporation\Hawkrad, to make contact with such an effective Hilbert-space description.  Then, we will describe models for unitary but nonlocal evolution.  Such models at the least may guide understanding of the constraints on unitary scenarios, but might also guide deeper understanding of the principles of quantum gravity.

\newsec{Hawking evaporation}

We take rotation to be an inessential complication and consider a Schwarzschild metric,
\eqn\Sch{ds^2=-f(r)dt^2+{dr^2\over f(r)} + r^2 d\Omega^2\ .}
This form is valid in general spacetime dimension $D$, although one may particularly focus on $D=4$ where $f(r)= 1-2M/r$.  Hawking radiation is conveniently analyzed by introducing tortoise coordinates, via $dr^*/dr = 1/f(r)$.  Near infinity, $r^*\approx r$, but as the horizon $r=R$  is approached, $r^*\rightarrow -\infty$.  Consider a free field $\Phi$, with general spin.  Solutions of the corresponding wave equation may be written as a sum of terms
\eqn\wavesoln{\Phi_{J} = {u_J(r,t)\over r^{D/2-1}} Y_J(\Omega)\ ,}
where $J$ labels angular momentum and $Y_J$ is an appropriate spherical harmonic.  In terms of $(r^*,t)$, $u_J$ satisfies a two-dimensional wave equation with  effective potential that vanishes at $r^*=\pm\infty$, and has a maximum $\sim J^2/R^2$.  Further details appear in, {\it e.g.}, \refs{\SGTrieste}.

The quantum field outside the BH may be expanded in modes $\Phi_{J,\omega}$, of energy $\omega$, and creation/annihilation operators, as $\Phi=\sum_J\int d\omega (\Phi_{J,\omega} a_{J,\omega}+h.c)$.  As reviewed in \SGTrieste, the Hawking state is found by determining the transformation between the coordinate $x^-=t-r^*$, in which the ``out" vacuum is naturally defined, and appropriate coordinates for defining an ``in" vacuum.  One may also choose corresponding modes\refs{\GiNe,\SGTrieste} ${\hat \Phi}_{J,\omega}$ inside the horizon, with mode operators ${\hat a}_{J,\omega}$.  Then, the ``in" vacuum evolves to the Hawking state, which takes the form
\eqn\hawkstate{|\psi\rangle = c \sum_{\{n_{J,\omega}\}} e^{-{H/2T} } |\{{\hat n}_{J,\omega}\}\rangle  |\{n_{J,\omega}\}\rangle\ .}
Here $n_{J,\omega}$ are occupation numbers, $T=(D-3)/4\pi R$ is the temperature,
\eqn\hamil{H=\sum_J\int d\omega\, \omega\, n_{J,\omega}}
is the hamiltonian in the Fock basis we have chosen, and  $c$ is a constant.  Tracing out the internal states gives a thermal density matrix.

In order to match to an effective Hilbert space description, it is useful to choose a wavepacket basis.  For example, a simple set of complete, orthonormal states investigated in \refs{\Hawkrad,\GiNe} is, with integer $k$ and $n$, 
\eqn\wavebas{u_{J,k,n} = \epsilon^{-1/2} \int_{k\epsilon}^{(k+1)\epsilon} d\omega e^{2\pi i \omega n/\epsilon} u_{J,\omega}\ .}
These have frequency $\omega\simeq k\epsilon$ and are localized about $x^-=2\pi n/\epsilon$ with width $1/\epsilon$.  The variable $\epsilon$ is an arbitrary choice; it is convenient to take $\epsilon\roughly> 1/R$.  Other, smoother, wavepacket bases may be chosen, but \wavebas\ is simple and intuitive.  The hamiltonian $H$ may be easily reexpressed in such a basis.

We can now describe an effective Hilbert-space model for evolution governed by $H$, following discussion of \refs{\Page\QBHB\HaPr-\Braunstein,\Mathurrev,\Rozali}.  It is convenient to do so by describing the modes and their state in terms of a time slicing.  One such choice is a nice slicing, as described in \refs{\LPSTU,\QBHB,\Mathurrev,\NLvC}; in the particular realization of \NLvC\ the slice $S_t$ asymptotes to the constant $t$ slice at $r=\infty$; inside the horizon and in the far past it asymptotes to $r=r_c$.  In the static BH geometry slices at different $t$ are just $t$-translates of this slice; with decreasing BH mass minor adjustments to such slices are needed.

On such slices, evolution of the state \hawkstate\ in a basis such as \wavebas\ may be pictured as follows.  First, excited quanta at $r\gg R$ simply evolve outwards.  Quanta at $r<R$ evolve inwards; in the nice slice description they then freeze at $r=r_c$, though evolution would continue in a ``natural" slicing\NLvC\ approaching $r=0$.  Finally, the evolution produces paired quanta from $r=R$.  When their wavelength on the slice is $\ll \omega^{-1}$, the pair is nearly indistinguishable from vacuum, {\it e.g} by gravitational scattering\BHIUN, but when their wavelength reaches its asymptotic value $\sim \omega^{-1}$, the quanta separate from each other and the horizon and travel into/out of the BH.  

Excited quanta typically have $\omega\sim T\sim 1/R$ and are emitted every time $\delta t\sim R$.  Thus, in a simple model (see {\it e.g.} \Mathurrev) we consider one species of particle and ignore spin, set $\epsilon=1/R$, and only keep the states with $k=1$, so $\omega=1/R$, and with occupation number zero or one.  Then, evolution can be described in time steps of size $\delta t$, as follows.  First, it suffices to describe evolution of a basis of states for the combined black hole and external Hilbert spaces, $\calhh\otimes\calh$.  Such a basis can be written in terms of states $|\ahat\rangle|a\rangle$, where for example the states $|\ahat\rangle$ and $|a\rangle$ are bases for $\calhh$ and $\calh$.   The general combined state (and thus general initial state) takes the form $\sum_{\ahat a} c_{\ahat a} |\ahat\rangle|a\rangle$.  We describe $\ahats$ and $\as$ in terms of strings of qubits, and model evolution for $\delta t$ as
\eqn\paircreate{\ahats\as \rightarrow \uhat \ahats\otimes {1\over \rtt}\left(\zhats \zs+ \ohats\os\right)\otimes U \as\ .}
Here, $U$ and $\uhat$ are unitary operators that we may think of as describing evolution given by \hamil\ of the external wavepackets away from the black hole, and of the internal wavepackets, respectively.  Different slicings/mode descriptions yield different $U,\uhat$, but equivalent evolution.  In addition to this evolution, an entangled pair of quanta is produced from the horizon.

When backreaction is included, $M$ decreases an amount $\sim 1/R$ in each time step, and so shrinks to near zero at a time $T_{evap}\sim RS_{BH}$.  The late-time external state, with $\calo(S_{BH})$ quanta, is found by tracing over internal states in \paircreate (or in \hawkstate)
\eqn\rhoout{\rho = {\rm Tr}_{\calhh}\left( |\psi\rangle \langle \psi |\right)\ ,}
and  is mixed, with entropy $S=-{\rm Tr}(\rho\log\rho) \sim S_{BH}$. Barring a remnant scenario (which is argued on other grounds\refs{\wabhip,\Sussrem} to be unphysical) the black hole and internal Hilbert space disappear, leaving this mixed state.  This is Hawking's basic argument\refs{\Hawkunc} for information loss -- which ultimately conflicts with energy conservation\refs{\BPS}.

\newsec{Models of unitary evolution}

The basic conflict between QM/energy conservation, locality, and Lorentz/diffe\-omorph\-ism invariance that arises in the previous scenario has been called the ``information paradox."  It was argued in \refs{\QBHB,\NLvC} (see also \refs{\GiLi,\LQGST,\BHIUN}) that this is not a true paradox, in that we don't have a sharp derivation of the state \hawkstate\ and density matrix \rhoout\ in a perturbative framework that takes into account backreaction.  However, as discussed in \refs{\GiLi,\LQGST,\BHIUN,\QBHB,\NLvC}, it seems evident that some amount of nonlocality with respect to the semiclassical picture is needed to avoid the essential argument for lost information.  This section will model a kind of nonlocal evolution apparently needed for unitarity, in the effective Hilbert-space approach.

As one guide, we begin with the expectation that the internal Hilbert space of the BH shrinks, and contains no information when $M\rightarrow0$. A candidate parameterization for $N(M) = \log({\rm dim} \calhh)$ is  \ndof, but one could also consider other functions decreasing to zero.  This contrasts with the evolution \paircreate, in which $N(M)$ increases by one in each step (being trivially incautious in distinguishing bits and nats).

We can model unitary evolution, in which $N(M)$ decreases by one at each step, in different modifications of \paircreate. One is, separating off the leftmost qubit and choosing a basis element $\ahats$ for the remaining internal Hilbert space,
\eqn\evolone{\zhats \ahats \as \rightarrow \uhat\ahats \otimes \zhats \zs \otimes U \as\quad ,\quad \ohats \ahats \as \rightarrow \uhat' \ahats \otimes \zhats \os \otimes U\as\ ,}
which is effectively unitary. Moreover, if the rightmost internal qubit is always in a definite state, here $\zhats$, it can be forgotten or ``erased" with impunity.  Thus, internal information is transmitted to the external state, and $N(M)$ decreases by one.\foot{Note also a relation with the final-state picture of \refs{\HoMa}; if a given internal qubit first transitions to a given canonical state, this qubit may then be unitarily forgotten.  See also \Rozali.}   

In \evolone\ information is relayed from the leftmost of $N(M)$ qubits.  In case of a general unitary transformation $\uhat$ in the preceding time step, this has no invariant meaning.  One may take different models for $\uhat$ (also with infalling matter -- see below).  One is simply $\uhat=1$, in which case the leftmost qubit from evolution \paircreate\ would not be occupied until a time $t\sim R S_{BH}$, and so information return only begins at this time.  Another is a random unitary\refs{\Page,\Braunstein,\BaFitherm}.  This can be thought of as a model of fast-scrambling\refs{\HaPr,\SeSu}.  Different models may be distinguished\NLvC\ by this retention time $T_r$, describing how long it takes the infalling quanta to mix with the escaping qubits.  Note that, even with rapid mixing, the effect of a given infalling quantum initially has tiny effect on the outgoing state, but later in evolution discussion of \refs{\HaPr} can apply.
Note also that in the semiclassical/Hawking picture \paircreate, $T_r=\infty$, and in this sense the longest possible\refs{\Page} $T_r$, $\sim R S_{BH}(M)$, is most conservative.

Eq.~\evolone\ represents a big departure from \paircreate, particularly if present in early evolution of the BH.  One may consider alternatives with smaller departures.  One is
\eqn\evoltwo{\eqalign{\zhats\zhats \ahats \as &\rightarrow \uhat \ahats\otimes {1\over \rtt}\left(\zhats \zs+ \ohats\os\right)\otimes U \as\quad,\quad  \zhats\ohats \ahats \as \rightarrow \uhat \ahats\otimes\zhats\os\otimes U\as\cr \ohats \zhats \ahats \as  &\rightarrow \uhat \ahats\otimes\ohats\zs\otimes U\as\quad,\quad \ohats\ohats \ahats \as \rightarrow \uhat \ahats\otimes {1\over \rtt}\left(\zhats \zs- \ohats\os\right)\otimes U \as\ , }} 
which, with $\uhat=1$, does not alter the Hawking state until $\sim T_{evap}\sim R S_{BH}$.   Other simple generalizations clearly exist, including using other pairs.\foot{Note that a version where one internal qubit is imprinted in a pair does not relay a net bit of information \refs{\Mathurrev,\GiSh}.} Yet another alternative (also generalizable) uses other states, {\it e.g.} $|1'\rangle$, $|1''\rangle$, that have small amplitude for occupancy in the Hawking state \hawkstate:
%
\eqn\evolthree{ |\qhat_1\qhat_2\rangle \ahats \as \rightarrow \uhat \ahats\otimes {1\over \rtt}\left(\zhats \zs+ \ohats\os\right)\otimes|\zhat'\zhat''\rangle|q_1'q_2''\rangle\otimes U \as\ ,}
where $q_1,q_2=$ 0 or 1, independently.

The unitary evolution laws \evolone-\evolthree\ are clearly nonlocal with respect to the semiclassical geometry of the BH.  One may worry that there is a more serious objection, namely that anything but the evolution \paircreate\ (or \hawkstate) produces a state that an infalling observer sees as very singular, giving a large departure from expected BH behavior.  (The pairing in \hawkstate\ yields cancellations\refs{\BHIUN} between the contributions of the quanta, interacting with an infalling observer.)

Here, allowing such modest nonlocality can be an asset.  In the given basis, we can think of the state of paired quanta near the horizon as being of the form $ {1\over \rtt}\left(\zhats \zs+ \ohats\os\right)$ until the time these quanta have wavelength $\sim R$, on which time scale the transitions like  in \evolone-\evolthree\ occur.  So, departures from the Hawking state only arise for quanta with wavelengths of order the horizon size.  One may imagine these departures as arising from some new nonperturbative, perhaps collective, effect of the quantum BH.
This evolution results in modifications to $\calo(1)$ quanta per time $R$,  that an infalling observer sees as having energy $1/R$ -- apparently for a large black hole, a very small effect.  Such effects seem particularly small if they modify, for example, graviton states --  these would be essentially indetectable for an infalling observer.

Modifications to LQFT evolution like those described in \NLvC\ and modeled here are apparently the minimal required to extract information from a BH.  Note that $U$ in \evolone-\evolthree\ is taken to describe evolution as in LQFT.  In general evolution inside is only expected to be governed by LQFT for a typical time $\sim R$ it takes a quanta to fall into the strong curvature domain; after that we have few constraints on $\uhat$.  One can artificially freeze this evolution in a nice slicing.  However, this becomes an extreme construction when extended over long times, and in particular a perturbative LQFT quantization on nice slices has been argued to be problematic\refs{\QBHB,\GiLi,\BHIUN}.  This motivates the departure from LQFT of the types \evolone-\evolthree\ (and generalizations), which depart from LQFT in the $\uhat$-evolution for qubits deep inside the BH and being relayed outside the BH, and for the qubits in a region of size $\sim R$ receiving the information.  Note also that spacelike communication in {\it flat} space can be related to acausal communication, by a boost.  But, the BH geometry breaks the boost symmetry, fixing a frame with respect to which the evolution can be causal\refs{\NLvC}.

Evolution of types \evolone-\evolthree\ represent different pictures with the common feature that a net one qubit is relayed from inside to outside at each step in time.  Eq.~\evolone\ is a big change to \hawkstate, in which quanta are not produced paired with internal Hawking excitations.  It is a simple model for evolution if true Hawking radiation is not present, such as could occur in a general massive remnant scenario\refs{\BHMR} or special cases of it such as fuzzballs\refs{\fuzz}.  Eq.~\evoltwo, with $\uhat=1$, postpones significant modification until $t\sim R S_{BH}$ when the BH has shrunk appreciably. After this time, it departs from the Hawking state, but has some of its features, in particular the same average outward flux.  And, \evolthree\ yields extra outward flux, in addition to Hawking evaporation; again with $\uhat=1$ this may be postponed to $t\sim R S_{BH}$.  Straightforward generalizations of \evolone-\evolthree\ clearly exist.

\newsec{Refinements and enhancements}

Various refinements are possible, to bring our models closer to a complete description.  First, one can clearly generalize evolution of types \evolone-\evolthree\ to incorporate the many different modes in \hawkstate.  There is corresponding flexibility in which modes \evolone-\evolthree\ imprint the information, though we may assume they act on modes in a region of size $\calo(R)$ near the BH and only on modes close to their asymptotic wavelengths ({\it i.e.} not highly blueshifted).  One may also choose different evolutions corresponding to different rates of reduction of the size $N(M)$ of the BH Hilbert space, though \ndof\ seems natural.  

One may also incorporate infalling matter, {\it e.g.} in modes with wavelength $\ll R$.  The LQFT description of this is a straightforward extension of the discussion in sec.~2, and we might expect its evolution by $\uhat$ until it reaches strong curvature to be approximately that of LQFT.\foot{One might also choose slicings staying outside the BH to give holographic\refs{\thooholo,\sussholo} $\uhat$.}  After this, again there are few constraints on the subsequent $\uhat$ evolution on such modes. One attractive possibility is that a particle of wavelength $\ll R$ is ``broken up" into a collection of the internal modes described above, before the information is relayed (and unoccupied bits erased) in transformations like \evolone-\evolthree.  Note that since the entropy in infalling matter does not exceed $S_{BH}(M)$, we are not in danger of having more information encoded in the BH than can be relayed by these evolution rules.

When faced with a failure of classical mechanics in the atom, Bohr introduced a simple phenomenological model with new rules to capture the correct physical behavior; this led to the development of the  profound formalism of quantum mechanics.  If the world is quantum mechanical and yet locality holds to a good approximation, we can likewise consider simple models of unitary black hole evolution.  Due to the apparent failure of LQFT to describe this evolution, such a model no longer resides within that framework -- new principles are needed.  These models may clarify constraints on unitary evolution, and may be, as with Bohr's atom and quantum mechanics, guides to a deeper understanding of the more basic and complete non-local mechanics describing quantum gravity.

\bigskip\bigskip\centerline{{\bf Acknowledgments}}\nobreak

I thank S. Mathur, D. Page, and Y. Shi for useful conversations.  This work  was supported in part by the Department of Energy under Contract DE-FG02-91ER40618 and by grant FQXi-RFP3-1008 from the Foundational Questions Institute (FQXi)/Silicon Valley Community Foundation.

\listrefs
\end